# High efficiency symmetric beam splitter for cold atoms with a standing wave light pulse sequence

Saijun Wu[1,2], Ying-Ju Wang[3], Quentin Diot[3], Mara Prentiss[1]

1. Department of Physics and Center of Ultra Cold atoms, Harvard University, Cambridge, MA, 02138

2. Division of Engineering and Applied Science, Harvard University, Cambridge, MA, 02138

3. Department of Physics, University of Colorado, and JILA, National

Institute of Standards and Technology and University of Colorado, Boulder,

Colorado 80309-0440, USA

## Abstract:

In a recent experiment [1], it was observed that a sequence of two standing wave square pulses can split a BEC at rest into +/- $2\hbar\mathbf{k}$ diffraction orders with almost 100% efficiency. By truncating the Raman-Nath equations to a 2-state model, we provide an intuitive picture that explains this double square pulse beamsplitter scheme. We further show it is possible to optimize a standingwave multi square pulse sequence to efficiently diffract an atom at rest to symmetric superposition of +/- $2n\hbar\mathbf{k}$ diffraction order with n>1. The approach is considered to be qualitatively different from the traditional light pulse schemes in the Bragg or the Raman-Nath region, and can be extended to more complex atomic optical elements that produce various tailored output momentum states from a cold atom source.

## Introduction

For atom interferometry it is useful to have large angle beamsplitters that put the input atom wavepackets into superposition of two narrow momenta distributions with large momentum separations. The narrow distributions are required to obtain good fringe contrast. The large momentum separation not only helps to spatially resolve the two arms of the interferometer but also reduces the effects of stray fields. The efforts toward achieving a matterwave large angle beamsplitter with light pulses may be divided into two categories, depending on whether or not the external motions of the atoms are separable from the internal states. In the first approach, the atoms have to be in the superposition of two different internal states that are associated with two different external motional states. A predetermined number of photon momenta can be transferred to atoms by successively inducing the transitions between the two states while alternatively changing the direction of the light fields [2-4]. In the second approach, the atoms stay in a single (adiabatic) internal state throughout the interaction, and the motion of atoms can be described by a scalar matterwave in a periodic light shift potential [5]. To narrow the output momentum transfer spreading in the second approach, either the light shift potential is carefully designed [6, 7], or the initial momentum of the atom is controlled to meet the Bragg resonance [8-10].

For interferometry purposes, if the atoms leaving the beamsplitter in different momentum states are also in different internal states, the energy difference between these internal states can contribute to the phase error in an interferometer, since the energy difference is usually sensitive to the stray fields. With multi-photon Bragg scattering, a single frequency standing wave light field can cleanly split the cold atoms into two diffraction orders with large momentum separation, while throughout the process the atoms are in a single (adiabatic) internal state. The authors in [11] suggest that with the development of cold atom technology, the beamsplitter based on the Bragg scattering could be a promising alternative to those work with multi atomic internal levels.

Manipulation of matterwave with an off-resonant standingwave light pulse is a subject that has been extensively studied [12]. A Kapitza-Dirac pulse can be considered as complimentary to a Bragg pulse, since the light pulse duration is so short that the motion of atom during the pulse is ignorable (Raman-Nath, or thin lens limit). A Kapitza-Dirac pulse shares the common advantage of light pulse beamsplitter schemes work in the Raman-Nath limit (input velocity insensitive/achromatic), thus is advantageous if a collimated atom source is not available [13]. However, the multi-momenta output limits the achievable fringe contrast in the scheme. On the other hand, if a collimated atom source is available, the light pulse may be designed to populate only certain output diffraction orders that are relevant for a particular experimental purpose [14].

The work in this paper is motivated by a recent experimental discovery [1]: A sequence of two square shaped standing wave light pulses can split a BEC at rest into +/- $2\hbar\mathbf{k}$ diffraction orders with almost 100% efficiency. Different from a Bragg pulse, the double square pulse transfers not only momentum, but also energy to the atoms. On the other hand, the scheme transfers momentum to the atom in a highly controlled manner, in contrast to a Kapitza-Dirac pulse.

The first part of the paper is direct connected to the experimental discovery. By mapping the dynamics of the matter wave onto a two-state system, we explain how a double square pulse is able to diffract atom at rest to symmetric n=+/–1 diffraction orders with almost 100% efficiency. Numerical simulations are used to support the analysis. The analysis is also shown to be in quantitative agreement with the experimental discovery [1].

In the second part of the paper, we describe matter wave dynamics under a sequence of square shaped light pulses from a multi-path interference point of view. We then numerically optimize the double square pulses as $2n\hbar\mathbf{k}$ beamsplitters for n from 1 to 6 and show the diffraction efficiency to a desired diffraction order can be significantly higher than that achievable by a Kapitza-Dirac pulse. We expect the experimental demonstration of these $2n\hbar\mathbf{k}$ beamsplitters (with n>1) in the near future.

We note that the related beamsplitter schemes with standing wave light field are developed in [15-17]. For example, it is shown in [16] that in the "pump-probe" geometry, the Bragg scattering by two opposite moving standing waves can be continued n times to put atoms into a symmetric superposition of $+/-2n\hbar\mathbf{k}$ diffraction order. However, in [15~17] resonance and adiabacity are the key elements to narrow the diffraction to a specific order, while here the same goal is achieved by properly timing the square pulses to control the multi-path interferences. We consider the work described here as a novel strategy in the standing wave control of atomic motion.

# I. $2\hbar k$ symmetric beamsplitter

We begin our discussion with a physical model based on a two-level atom interacting with a far-off resonant light field so that the excited state of the atom is adiabatically eliminated. The atom is thus described by a free scalar matterwave and the sinusoidal AC stark shift is the only interaction introduced by the classical standing wave light field. The motion of the atom can be described by the coupled Raman-Nath equations (RNE). We cut off the RNE to include only the n=0 and n=+/–1 diffraction orders. The model provides an intuitive picture that explains why a sequence of two square pulses can convert an atom at rest to the symmetric superposition of +/– $2\hbar k$ diffraction orders with almost 100% efficiency. We then integrate the full RNE numerically to confirm the conclusion from the simple model.

## I.1 From the Raman-Nath Equations to a 2-level system

For atom in a standing wave light shift potential, the Schrödinger's equation is given by:

$$i\dot{\psi}(x,t) = [-\frac{\hbar}{2m}\frac{d^2}{dx^2} + \Omega(t)Cos(2k_0 x)]\psi(x,t) \tag{1}$$

Here $\Omega(t)$ describes the amplitude of the light shift potential and $k_0$ is the wavevector of the light field. Equation (1) is restricted to the situation where the nodes of the standing wave are fixed in space, which corresponds to the familiar situation where the standing wave is formed by reflecting back a traveling wave with a fixed mirror.

By expanding the wavefunction in the (empty lattice) Bloch basis,

i.e: $\psi(x,t) = \int dk \sum_n C_{2n}(k,t) e^{i(2nk_0+k)x}$, and substitution in Eq. (1) we obtain the Coupled

Raman-Nath Equations:

$$i\dot{C}_{2n}(k,t) = \frac{\hbar}{2m}(2nk_0+k)^2 C_{2n}(k,t) + \frac{\Omega(t)}{2}[C_{2n-2}(k,t) + C_{2n+2}(k,t)] \qquad (2)$$

First let's consider the atom with initial conditions $C_{2n}(k,0) = \delta_{n,0}f(k)$, with $|f(k)|^2$ giving a

narrow distribution around k=0 and $\Delta k \ll k_0$. Although (2) comprises a infinite set of

coupled equations, we may truncate them to effectively include only the lowest order

2N-1 equations, given that $\frac{\Omega(t)}{2} \ll (2N)^2 \frac{\hbar k_0^2}{2m}$. Here we choose N=2 and restrict the coupling constant $\Omega(t) \ll 32 \frac{\hbar k_0^2}{2m} = 32\omega_r$. In this case equation (2) includes only velocity

classes with n=-1, 0, 1. With a unitary transformation on $C_{2n}(k,t)$ to cancel out the kinetic

energy associated with k, we have:

$$i\dot{C}_0(k,t) = \frac{\Omega(t)}{2}[C_{-2}(k,t) + C_{+2}(k,t)]$$

$$i\dot{C}_2(k,t) = 4\omega_r(1+\frac{k}{k_0})C_2 + \frac{\Omega(t)}{2}C_0(k,t)$$

$$i\dot{C}_{-2}(k,t) = 4\omega_r(1-\frac{k}{k_0})C_{-2} + \frac{\Omega(t)}{2}C_0(k,t) \qquad (3)$$

It is helpful to rewrite (3) into the form:

$$i\dot{C}_0(k,t) = \frac{\Omega(t)}{\sqrt{2}}C_+$$

$$i\dot{C}_+(k,t) = 4\omega_r C_+ + \frac{\Omega(t)}{\sqrt{2}}C_0 + 4\omega_r \frac{k}{k_0}C_-$$

$$i\dot{C}_-(k,t) = 4\omega_r C_- + 4\omega_r \frac{k}{k_0}C_+ \qquad (4)$$

With $C_+ = \frac{1}{\sqrt{2}}(C_2 + C_{-2})$ and $C_- = \frac{1}{\sqrt{2}}(C_2 - C_{-2})$.

If $\frac{k}{k_0}$ terms are ignored in (4), only $C_+$ is coupled to $C_0$, and $C_-$ becomes the amplitude of a dark state. Thus the dynamics of the light-atom interaction reduces to a 2-state system.

Choosing a different rotating frame, we obtain,

$$i\dot{C}_0 = -2\omega_r C_0 + \frac{\Omega(t)}{\sqrt{2}} C_+$$

$$i\dot{C}_+ = \frac{\Omega(t)}{\sqrt{2}} C_0 + 2\omega_r C_+ \quad\quad\quad (5)$$

In the next section, we will discuss how to transfer the population from $C_0=1$ to $C_+=1$ by a double pulse sequence while keeping $\Omega(t) << 32\omega_r$.

## I.2 Optimization of the 2-pulse for 2-photon transition

In nuclear magnetic resonance experiments one often inverts the spin of a nucleus with a sequence of weak rf pulses rather than a single strong pulse to avoid complications from other dimensions in the Hilbert space. Here we have similar problem: we want to invert the population of $C_0$ and $C_+$ in (5) while keeping $\Omega(t) << 32\omega_r$ so that the cut-off we made to give (5) is justified. The simplest way to do this is to use a compound pulse composed of two square pulses. The 1$^{st}$ square pulse can put the system onto an equal superposition of $C_0$ and $C_+$. After the re-phasing of $C_0$ and $C_+$ during the free evolution, the population can be completely transferred from $|C_0|^2=1$ to $|C_+|^2=1$ by the 2$^{nd}$ pulse. This is explained in Fig.1 where we introduce the Bloch sphere to describe the evolution of the matterwave states. This simple 2-state model gives the optimal parameters of the double square pulse for the beamsplitting. They are: (a) the square pulses should be with amplitude $\Omega_m = 2\sqrt{2}\omega_r$, (b) the pulse durations for the square pulses should be chosen as $\tau_1 = (2n_1+1)\frac{\pi}{4\sqrt{2}\omega_r}$, and (c) the interval between the two pulses should be chosen as $\tau_2 = (2n_2+1)\frac{\pi}{4\omega_r}$. Here $n_1$ and $n_2$ are two integers. Notice the optimal parameters only require the maximum pulse strength $\Omega_m$ to be $2\sqrt{2}\omega_r$, which is fairly compatible with the assumption $\Omega(t) << 32\omega_r$.

We numerically integrate equation (2) at k=0 with initial condition $C_{2n}(k=0,0)=\delta_{n,0}$. $\Omega(t)$ is set to be the double square pulse with the amplitude $\Omega_m = 2\sqrt{2}\omega_r$, and with different $\tau_1$ and $\tau_2$. The output $|C_2(k=0,2\tau_1+\tau_2)|^2$ is plotted as a function of $\tau_1$ and $\tau_2$ in fig. 2. Also included in fig. 2 is the numerical integration of the 2-state model. We see that the double square pulse process here can almost quantitatively be described by the 2-state model when $\tau_1$ is small: A double pulse with the parameters prescribed by the simple model almost completely inverts the population. In the full model, however, the optimal values on the $\tau_1-\tau_2$ plane are periodically distributed double peaks instead of single peaks. The correction is due to the finite coupling to higher diffraction orders during $\tau_1$. The numerical simulation indicates that the optimal conversion efficiency is very close to unity with error only of order $O(10^{-4})$.

### I.3 The initial velocity dependence of the $2\hbar k$ symmetric beamsplitter

In this section we consider the influence of small initial velocity for the performance of double square pulse that is optimized for k=0 velocity class. The initial kinetic energy term in equation (4) introduces the coupling between the symmetric state labeled by $C_+$ and anti-symmetric state labeled by $C_-$. For k<<$k_0$, the amplitude at $C_-$ after the double pulse is of order $4\omega_r \frac{k}{k_0}\frac{1}{2}(2\tau_1+\tau_2) \sim \pi\frac{k}{k_0}(\frac{2n_1+1}{\sqrt{2}}+\frac{2n_2+1}{2})$. To ensure $|C_-(k,2\tau_1+\tau_2)|^2<<1$, the initial momentum of the atom should satisfy

$$\frac{k}{k_0} << \frac{1}{(\frac{2n_1+1}{\sqrt{2}} + \frac{2n_2+1}{2})\pi} \tag{6}$$

Obviously, for beamsplitting purposes, both $n_1$ and $n_2$ should be chosen as zeros such that the effect of the initial velocity spreading is minimized. This requires the initial atom sample to be cooled to sub-recoil temperatures.

We have also numerically integrated equation (4) to investigate the velocity dependence of the beamsplitter pulse. The results agree with our analysis, some of which are summarized in the first graph of Fig. 4.

## II. Extension to $2n\hbar k$ symmetric beamsplitter with n>1

From part I we see it is possible to symmetrically split atom at rest to $+/- 2\hbar k$ resulting in a $4\hbar k$ momentum separation. In this section, we aim to extend the idea and explore the possibility of symmetric splitting into higher diffraction orders. Unfortunately, the simple 2-state picture in the last section cannot be applied here since more than 3 motional states should be included in the discussion. In this section, we first consider the motion of atom subject to multi square pulses from a multi-path interferometric point of view. Then we numerically optimize the double square pulses for $2n\hbar \mathbf{k}$ splitting with n from 1 to 6, which are shown to be with efficiencies significantly higher than those achievable with a Kapitza-Dirac pulse.

## II.1 Matter wave dynamics in a square pulsed standing wave light field

In the physical model described by equation (1), for atom in a standing wave light field with slowly varying intensity $\Omega(t)$, the evolution of matterwave can be described in its eigenstate basis, e.g., the instantaneous Bloch basis associated with $\Omega(t)$. Sudden switching of light intensity to a different level is described by the diabatic projection of matter wave onto the new adiabatic basis.

This can be summarized into following two equations:

$$i\dot{C}_{2n}(k,t;\Omega) = E(k;\Omega)C_{2n}(k,t;\Omega) \tag{7}$$

and

$$C_{2n}(k,t^+;\Omega') = [\boldsymbol{A}(k;\Omega',\Omega)]_{nm}C_{2m}(k,t;\Omega) \tag{8}$$

Equation (7) describes the evolution of different adiabatic states during which light intensity is smooth (and in particular, constant). Different from (2), in (7) $C_{2n}$ as a function of $\Omega$ is the amplitude of the instantaneous Bloch states, and $E(k;\Omega)$ is the eigen frequency of the state. In equation (8), $C_{2n}$ from different Bloch basis are connected with the diabatic projection matrix $\boldsymbol{A}$.

Obviously, the properties of $E(k;\Omega)$ and $A(k;\Omega',\Omega)$ are essential in our analysis of pulse parameter designs [18]. For k=0 velocity class, if (7, 8) are truncated to include only the lowest 2N-1 equations, the system is reduced to an N level system with amplitudes labeled by $C_0$ and $C_{m+} \equiv \frac{1}{\sqrt{2}}(C_{2m} + C_{-2m})$, where 0<m<N. By adjusting the pulse duration, inter-pulse spacing, and pulse strength (which should be large enough to populate the desired order n, while smaller than a certain value so that the system can be approximated by a closed N level system), we expect the output of the "multi-path interferometer sequence" can be optimized for the population of a specific symmetric diffraction order.

In this paper we resort to numerical method in searching for an optimized multi-square pulse sequence for +/-2n$\hbar$k splitting. Limited by our computing resource, our simulation is restricted to double pulses with 4 adjustable parameters: $\Omega_m$ – common amplitude of the two pulses, $\tau_1$ – duration of the first pulse, $\tau_2$ – interval between the two pulses, and $\tau_3$ – duration of the second pulse (See fig. 3).

## II.2 Numerical optimization of +/-2n$\hbar$k splitting with double square pulses

Parallel to the discussion in part I, the optimization is taken out for k=0 velocity class. The range of the 4 parameters in optimization was chosen according to following criteria: to transfer population from zeroth order to nth order, $\Omega_m$ for 2n$\hbar$k splitting has to be comparable to the relative detuning $4n^2\omega_r$ but not too large such that the population diffuses away to even higher orders. The time parameters are restricted within the range $0 < \tau_1$, $\tau_2$, $\tau_3 < \dfrac{2\pi}{n\omega_r}$. The output $C_{n+} \equiv \dfrac{1}{\sqrt{2}}(C_{2n} + C_{-2n})$ is calculated by numerically integrating equation (1) and is optimized by $\tau_1$, $\tau_2$, $\tau_3$ at each sampling point of $\Omega_m$. The results from $2\hbar$k to $12\hbar$k are plotted with respective to $\Omega_m$ in fig 5. The dependence of the optimal population transfer on $\Omega_m$ is similar in all the numerical optimizations: after a critical strength of $\Omega_m$ the population at $C_{n+}$ approaches its optimal value, and then gradually becomes smaller as $\Omega_m$ continues to increase. From simulation we find the critical strength is roughly given by $\Omega_m = 2\sqrt{2} \times n^2 \omega_r$ for n=1 to 6 (See Fig. 3). Table 1 lists the optimal transfer efficiencies we get by this simple optimization scheme, which are significantly larger than those provided by a single Kapitza-Dirac pulse (the comparison is restricted to k=0 velocity class). We also varied the 4 parameters of the 2-pulse around the optimized value in the numerical simulations and looked at the sensitivity of the efficiency to these parameters. The results are also included in Table 1.

Having considered the case where the atom sample has no initial velocity, we will now consider the case where the atom sample has some initial velocity spread. In particular, we fix the double square pulse with parameters optimized for k=0 from the simulation and numerically integrate equation (2) that describe the velocity class with initial momentum $k_{off}$. The output momentum distribution is plotted in fig. 4 with respect to $k_{off}$. The results show that the optimized double square can address the atom distributions with momentum spread smaller than a pulse-related value $\Delta k$. This is also summarized in Table 1.

We emphasize that all the simulations in this work are based on the double square pulses with switching on and off in steps. The results in this paper remain valid if the switching time is finite but much smaller than the associated external motional time scale, e.g, $\frac{1}{4n^2 \omega_r}$, so that the diabaticity during the switching is satisfied. Typically the atomic recoil frequency is of order kHz, thus the switching time is well within current light pulse generation technology.

### III Discussion

The numerical optimizations in Part II have been restricted to the double square pulse configuration. However, we expect the optimal efficiency for $2n\hbar k$ splitting at large n be further improved with a pulse sequence composed of more than two square pulses. We also expect the optimization can be taken out analytically with the help from a deeper understanding of $E(k;\Omega)$ and $A(k;\Omega',\Omega)$ in equation (7, 8).

The $2n\hbar k$ symmetric beamsplitter discussed here is just one example that the motion of the atom wavefunction described by equation (1) can be controlled by adjusting the intensity of the standing wave light field $\Omega(t)$. In this work the optimization of the multi-square pulses has been considered for the diffraction of atoms to a single symmetric diffraction order. However, the optimization can also be taken out for a more general output momentum distribution, which may be relevant for a particular experimental purpose. For example, it can be shown that the signal of the "contrast interferometer" described in [14] is optimized if the atoms are in the *equal* superposition of 0 and symmetric +/-$2\hbar k$ diffraction orders during the free propagation. This momentum distribution can not be easily achieved using a single Kaptiza-Dirac or a Bragg pulse; however this distribution can be achieved with the multi-square pulse scheme here. Our numerical simulations show that the desired state can be produced by an optimal single square pulse with fidelity (the mode square of the inner product between the optimal state and the desired state) >99.6%. More generally, an equal superposition of 0 and the symmetric +/-$2n\hbar k$ diffraction orders with $n \geq 1$ has been numerically optimized with the double pulse configuration (with the procedures the same as described in Part II), with optimal fidelity found to be >99.99%, >99.8%, >99.0%, >92.5% for n=1, 2, 3, 4 respectively.

Throughout the discussion above we assume the nodes of the standing wave are fixed in space. If the relative phase between the two traveling waves (that forms the standing wave) is adjustable, equation (1) has to be modified as:

$$i\dot{\psi}(x,t) = [-\frac{\hbar}{2m}\nabla^2 + \Omega(t)Cos(2k_0x + \phi(t))]\psi(x,t) \tag{1'}$$

Thus two time-dependent functions $\Omega(t)$ and $\phi(t)$ can be adjusted to control the motion of atom. We note that the work described in [15-17] provides examples of controlling the motion of atoms by adjusting both $\Omega(t)$ and $\phi(t)$. Compared with the beamsplitter schemes proposed in [15-17], the scheme here only requires a standing wave with fixed nodes, which corresponds to the simplest possible standing wave setup (traveling wave + mirror). This is favorable in situations where simplicity and robustness are important.

In conclusion, we have described and numerically studied a new technique that efficiently splits atoms at rest into symmetric diffraction orders using double square pulses with optimized strength and intervals. We presented a simple and intuitive picture that explains the recent experimental discovery [1] and describes why the technique is effective for the 2-photon transition case. We further generally interpret the high efficiency of the beamsplitter scheme as a consequence of pre-designed multi-path interferences between different adiabatic diffraction orders. This work represents an extension of the field of coherent control for atoms, resulting in more efficient momentum transfer than single light pulse acting in the Raman-Nath regime. We expect a more complete description of optimal coherent control in system described by the Raman-Nath equations with square pulses. The beamsplitter scheme here can be extended to more complex atomic optical elements that produce various tailored output momentum states from a cold atom source.

## Acknowledgement:

We thank the inspiring discussions and the encouragements from Prof. Dana Anderson and Prof. Eric Cornell. We thank the critical reading and suggestions to the draft of this paper by Dr. Wilbert Rooijakkers. This work is supported by MURI and DARPA from DOD, NSF, ONR and U.S. Department of the Army, Agreement Number DAAD19-03-1-0106.

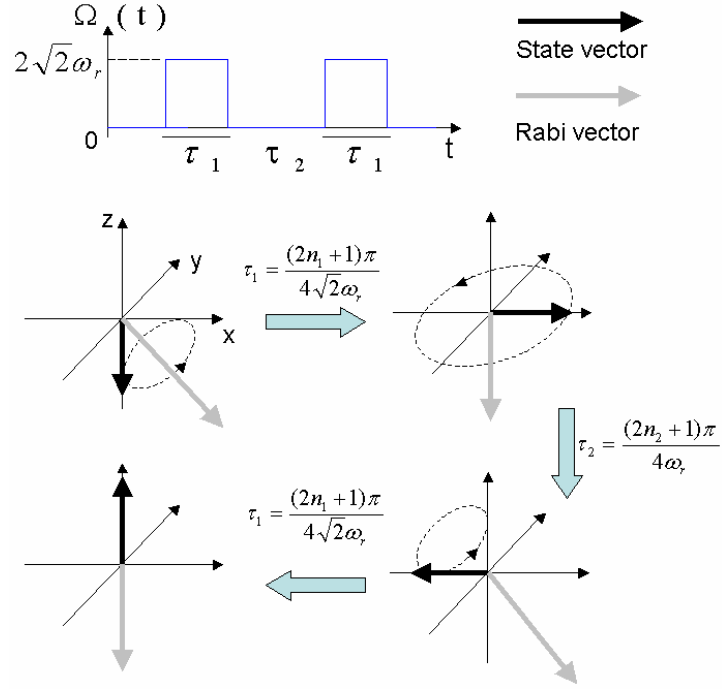

Fig. 1. The sequence of the 2-pulse (Top) and the Bloch sphere interpretation of the matterwave evolution (Bottom) according to equation (5) in the main text. Here $|\psi>=C_+|+>+C_0|0>$ $\equiv Cos[\frac{\theta}{2}]|+>+e^{i\phi}Sin[\frac{\theta}{2}]|0>$. With $|+>\equiv\frac{1}{\sqrt{2}}(|2>+|-2>)$ defined as the symmetric 2-photon diffraction state. Initially $\theta=\pi$ so that $C_+=0$. The four graphs on the bottom that are connected with three big arrows describe how the 2-pulse inverse the population and put $C_+=1$.

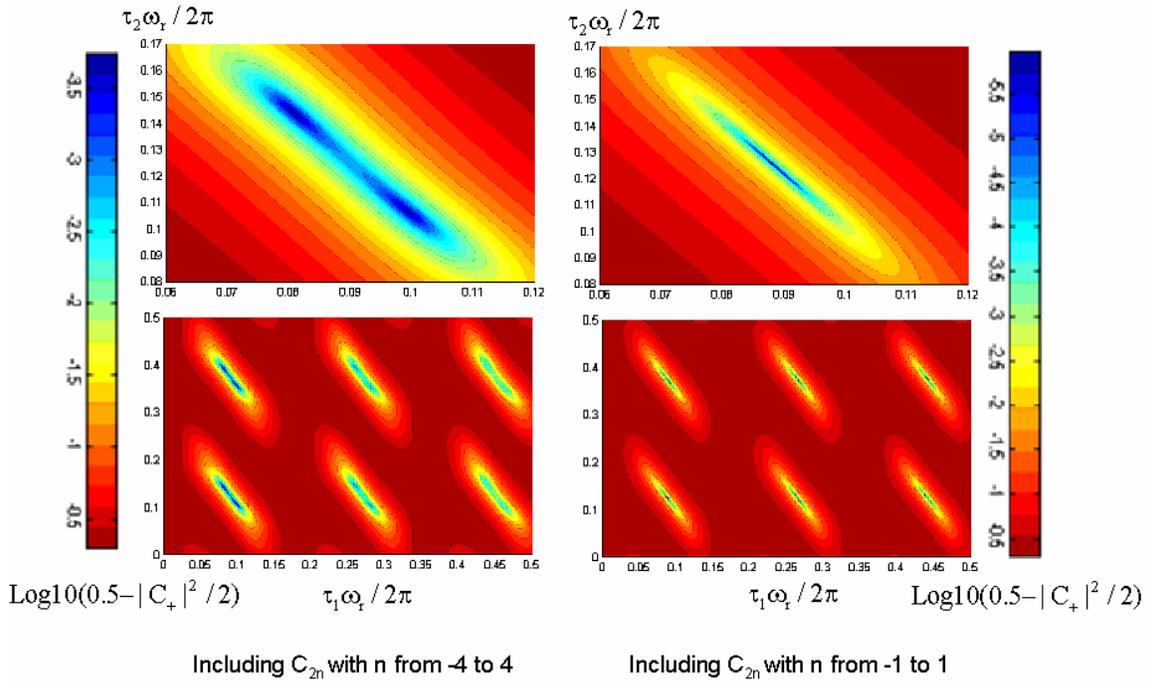

Including C$_{2n}$ with n from -4 to 4          Including C$_{2n}$ with n from -1 to 1

Fig. 2. (Color Online) Numerical results on the output population of the symmetric diffraction order C$_+$ as a function of the 2-pulse parameter $\tau_1$ and $\tau_2$. The colors are coded logarithmatically for $(1-|C_+|^2)/2$. Thus deeper blue indicate better diffraction efficiency. The two graphs on the left are calculated by numerically integrating the full equation (2) in the main text with k=0, while the two graphs on the right are the correspondent results based on the simple model with only 3 diffraction orders. The graphs on the top are respectively the zoom-in of the graphs on the bottom.

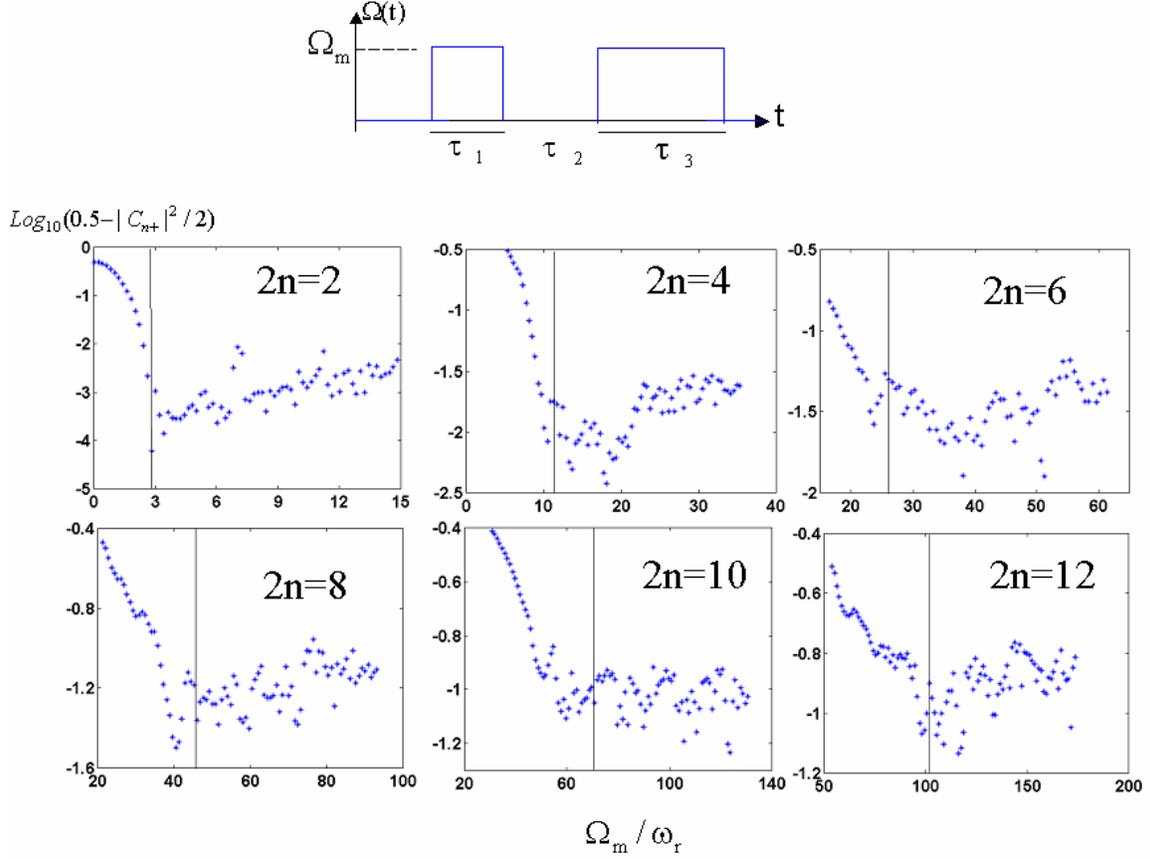

Fig. 3, Numerical optimization of the $2n\hbar k$ symmetric beamsplitter by numerically integrating equation (2) in the main text with k=0. Top: The shape of the 2-pulse with four adjustable parameters. Bottom: $|C_{n+}|^2 \equiv |\frac{1}{\sqrt{2}}(C_{2n}+C_{-2n})|^2$ is optimized at each point $\Omega_m$ by varying $\tau_1$, $\tau_2$ and $\tau_3$. The optimized (minimized) $(1-|C_{n+}|^2)/2$ is plotted logarithmatically with respect to $\Omega_m$. The results corresponds to n=1, 2, 3, 4, 5, 6 from Left to Right, Top to Bottom. In each graph a vertical line indicates the position $\Omega_m = 2\sqrt{2} \times n^2\omega_r$ which roughly gives the critical strength for the optimization. Notice at each $\Omega_m$ the optimized $|C_{n+}|^2$ is associated with a specific combination of $\tau_1$, $\tau_2$ and $\tau_3$, which is not shown.

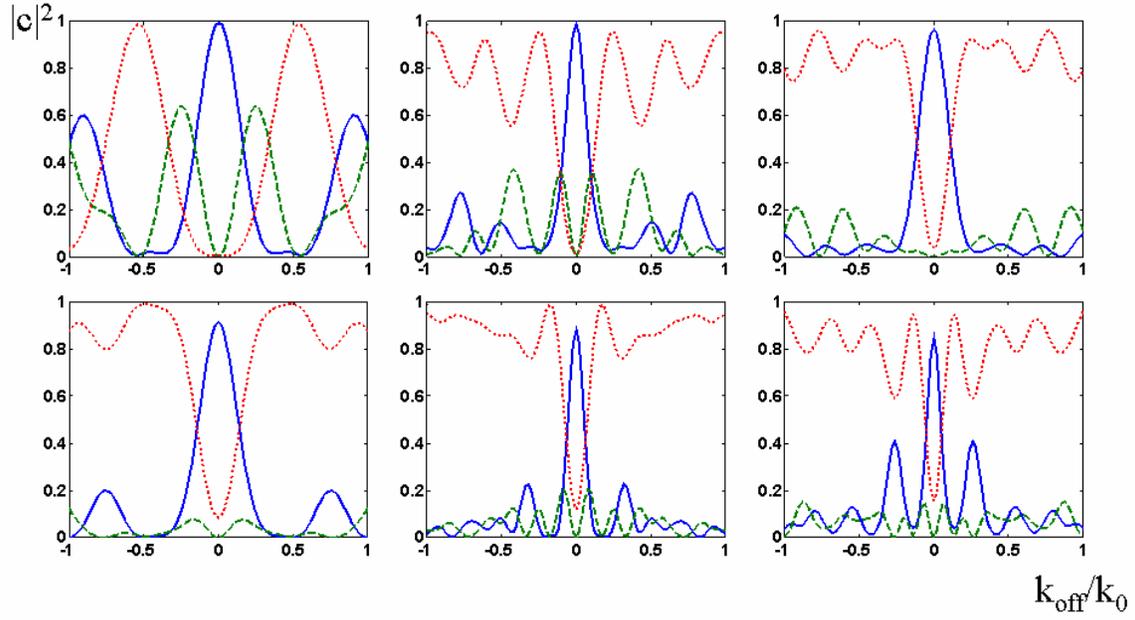

Fig. 4 Initial velocity dependence of the 2-pulse beamsplitter. The parameters of the 2-pulse are chosen from the optimized values in Table 1. The populations for $C_{n+} \equiv \frac{1}{\sqrt{2}}(C_{2n} + C_{-2n})$, $C_{n-} \equiv \frac{1}{\sqrt{2}}(C_{2n} - C_{-2n})$ are plotted as a function of the initial momentum $k_{off}$. With n=1, 2, 3, 4, 5, 6 from Left to Right, Top to Bottom. In each graph the solid line (-) gives $|C_{n+}|^2$, the dash line (- -) gives $|C_{n-}|^2$, the dotted line (..) gives population in other orders.

| $2n\hbar k$ | n=1 | n=2 | n=3 | n=4 | n=5 | n=6 |
|---|---|---|---|---|---|---|
| $\Omega_m / \omega_r$ | 2.83 | 13.4 | 33.9 | 57.6 | 124 | 109 |
| $\omega_r \tau_1 / 2\pi$ | 0.084 | 0.148 | 0.0290 | 0.0670 | 0.0090 | 0.0630 |
| $\omega_r \tau_2 / 2\pi$ | 0.143 | 0.112 | 0.0880 | 0.0450 | 0.0396 | 0.0315 |
| $\omega_r \tau_3 / 2\pi$ | 0.080 | 0.172 | 0.0310 | 0.0240 | 0.0846 | 0.1020 |
| $(|C_{n+}|^2)_{max}$ | **99.99%** | **99.1%** | **96.6%** | **91.7%** | **89.9%** | **85.5%** |
| $(|C_{n+}|^2)_>$ | 99.7% | 96% | 95.7% | 90% | 80% | 77.5% |
| $\dfrac{\Delta k}{k_0}$ | 0.31 | 0.16 | 0.22 | 0.28 | 0.12 | 0.11 |
| **Max[$J_n(\theta)^2$]** | **33.9%** | **23.7%** | **18.9%** | **16.0%** | **14.0%** | **12.5%** |

Table 1. Performance of the optimized 2-pulse for $2n\hbar k$ symmetric splitting. The optimized 4 parameters of the 2-pulse are listed in row 2-5. Row 6 lists the optimized transfer efficiency to the symmetric nth diffraction order $C_{n+} \equiv \dfrac{1}{\sqrt{2}}(C_{2n} + C_{-2n})$ with k=0. Row 7 gives the lower bound of the transfer efficiency when the last digits of the 4 parameters in row 2-5 vary by 1. In row 8 we list the momentum spreading tolerance with the transfer efficiency >50% of the optimal value for k=0 provided in row 6. In row 9 the maxima of the square of the nth order Bessel functions are listed in comparison with row 6, where the argument $\theta$ corresponds to the area of a Kapitza-Dirac pulse